\documentstyle[preprint,aps,version2]{revtex}
\begin{document}
\draft

\begin{title}
Proton-proton bremsstrahlung below and above pion-threshold:\\
the influence of the $\Delta$-isobar.
\end{title}

\author{F. de Jong$^{(1,2,3)}$, K. Nakayama$^{(1,2)}$,
T.-S. H. Lee$^{(4)}$}

\begin{instit}
$^{(1)}$Department of Physics and Astronomy, University of Georgia,
Athens, GA 30602\\
$^{(2)}$Institut f\"ur Kernphysik, Forschungszentrum J\"ulich,
52428 Germany\\
$^{(3)}$Kernfysisch Versneller Instituut, 9747 AA Groningen,
the Netherlands\\
$^{(4)}$Physics Division, Argonne National Laboratory, Argonne, IL 60439
\end{instit}

\begin{abstract}
The proton-proton bremsstrahlung is investigated
within a coupled-channel model with the $\Delta$ degree of freedom.
The model is consistent with the $NN$ scattering up to 1 GeV and the
$\gamma N\Delta$ vertex determined in the study of pion photoproduction
reactions.
It is found that the $\Delta$ excitation can significantly improve the
agreements with the $pp \rightarrow pp\gamma$ at $E_{lab}=280$ MeV.
Predictions at $E_{lab}=550$ and $800$ MeV are presented for future
experimental
tests.
\end{abstract}

\section{Introduction}

The Nucleon-Nucleon bremsstrahlung ($NN\gamma$) reaction has long been
considered as a tool to investigate the off-shell behaviour of
the $NN$ interaction.
With the availability of new experimental data of $pp\rightarrow pp\gamma$
reaction \cite{Michaelian,Przewoski},
theoretical interest has been revived recently
\cite{Fearing,Herrmann3,Herrmann1,Brown,Jetter,Katsogiannis}.
As a start, it is sufficient to consider only the one-nucleon current to
investigate this reaction since the leading interaction current
due to one-pion-exchange vanishes identically in the $pp \gamma$ reaction.
In a calculation taking into account the relativistic features of the
one-nucleon
current\cite{Herrmann3,Herrmann1}, the sensitivity of the
the $pp \rightarrow pp\gamma$ spin observables to
the $NN$ off-shell dynamics was demonstrated.
Although the calculated analyzing powers are
in general in good agreement with the data,
the cross-section results are more controversial.
In Ref.\ \cite{Michaelian} a normalization factor of 2/3 was proposed
to account for the discrepancy with the data.
However, this discrepancy might also indicate the importance of new
mechanisms not included in the calculations, thus omitting the need
of this rather arbitrary normalization factor.
In Ref.\ \cite{FdJ_ppg}, we have explored qualitatively the mechanisms
due to the $\Delta$ degree of freedom.
The main objective of this work is to present a consistent approach to
quantify our investigation. This is accomplished by extending the
coupled-channel formulation of $NN$ scattering developed
by Lee and Matsuyama \cite{Lee_1,Lee_2} to include the electromagnetic
interactions introduced by de Jong et al.\ \cite{FdJ_ppg}.

The effect of the $\Delta$ excitation on $NN\gamma$ has been
investigated in the past using very crude models.
Bohannon, Heller and Thompson \cite{Bohannon} derived the one-pion-exchange
$\Delta$ current by taking the static nucleon limit of the
pion photoproduction amplitude. The $\Delta$ width was neglected and hence
their model is limited to low energies.
They found that the $\Delta$ effect is negligibly small in $np$
bremsstrahlung. No calculation of $pp$ bremsstrahlung
based on their $\Delta$ current has been reported.
In contrast to the $np \gamma$ process, where the effect of the $\Delta$-isobar
is greatly reduced by cancellations between the different diagrams
involved due to iso-spin factors, the $\Delta$-effects may be significant in
$pp \gamma$ reactions.
Tiator et al.\  \cite{Tiator} evaluate
the contributions of the radiative $\Delta$-decay in Born-approximation and
add these incoherently to the nucleonic contributions which were calculated
using the Soft Photon Approximation (SPA).
Although Szyjewicz and Kamal \cite{Kamal} evaluated all
one-pion-exchange $\Delta$
contributions and add them coherently, their $pp\gamma$ calculation was
carried out using only the Born approximation.
In our previous work \cite{FdJ_ppg}, we added the single-scattering
$\Delta$-decay diagrams coherently to a state of the art calculation
of the nucleonic current contribution and found a significant effect.
In the present paper we go beyond this in two respects.
Firstly, we use off-shell T-matrix elements that are generated from a
meson-exchange coupled $NN\oplus N\Delta\oplus\pi NN$ model which is
constrained by the $NN$ scattering up to 1 GeV.
The quality of the phase-shift fit of this T-matrix is superior to the fit
of the T-matrix used in Ref.\ \cite{FdJ_ppg}.
Secondly, we also include the rescattering diagrams induced by the presence
of $\Delta$ degrees of freedom.
We follow the approach of Ref.\ \cite{Herrmann3} to account for the
relativistic features of the $N$ and $\Delta$ currents.
We will show that, even at energies below the pion threshold, the influence
of the $\Delta$  on the $pp\gamma$ reaction is considerable and
cannot be neglected in a quantitative comparison with the data.

In addition to giving a more complete description of $NN\gamma$ reaction
at low energies, our approach is appropriate for making predictions in the
intermediate energy region where
the pion production through the $\Delta$ excitation becomes crucial.
A good understanding of the $NN\gamma$ reaction at intermediate energies
is needed to describe the production of hard-photons in intermediate
energy heavy-ion collisions \cite{Nifen,Metag}.
In these complicated processes the photon provides a clean probe of
the reaction dynamics.
Furthermore, recent studies of dilepton production in proton-nucleus
collisions have shown that virtual $NN$ bremsstrahlung and $\Delta$-decay
are the dominant reaction mechanisms \cite{Naudet,Wolf}. The model presented
in this paper can be used as the starting point of a microscopic approach to
investigate both the photon and $\Delta$ productions in
relativistic heavy-ion collisions.

In section II, a coupled-channel formulation for $NN\gamma$ reaction will be
presented. The results are presented in section III. Section IV is devoted to
discussions of future developments.

\section{The Hamiltonian model of $NN$ and $NN\gamma$ interactions}

To investigate the effect of the $\Delta$ on the $NN\gamma$ reaction, it is
more transparent to employ a formulation including the $\Delta$
degree of freedom explicitly. In the first part of this
section, we will describe such an approach based on an
extension of the Hamiltonian model developed by Lee and Matsuyama
\cite{Lee_1,Lee_2} to
include the electromagnetic couplings introduced by de Jong et al.\
\cite{FdJ_ppg}.
We then describe how the procedures introduced in
Refs.\ \cite{Herrmann1,FdJ_ppg} are used to
include the "minimum relativity" in the $NN\gamma$ calculation.

The objective of the formulation of Refs.\ \cite{Lee_1,Lee_2}
was to obtain a consistent
description of $NN$ scattering from the low energy region
($ E_{lab} < 400$ MeV) to the intermediate energy region where the pion
production through the $\Delta$ excitation becomes important.
In contrast with the other $\pi NN$ models \cite{Gacilazo},
the formulation of Refs.\ \cite{Lee_1,Lee_2}
was designed to have a smooth transition to the usual nonrelativistic
potential model of $NN$ interaction.
This was achieved by using a substraction procedure to
define the $NN$ potential in the coupled $NN\oplus N\Delta\oplus \pi NN$ space
from a chosen $NN$ potential which fits the low energy $NN$ data.
The resulting model not only maintains the good fit to the low energy
$NN$ phase shifts,
but can also describe the $NN$ data up to about 1 GeV.
Starting with such a coupled-channel model will make the present study
significantly different from all of
the previous studies of the $\Delta$ effect on the $NN\gamma$ reaction.
In particular, we will be able to make realistic predictions
in the intermediate energy region where extensive data will soon become
available at COSY, the proton-cooler ring at the Forschungszentrum Juelich
(Germany).

In the formulation of Refs.\ \cite{Lee_1,Lee_2},
it is assumed that the Hamiltonian for $NN$ scattering can be
written in terms of three degrees of freedom: $N$, $\pi$, and $\Delta$.
In this work, we further assume that the $NN\gamma$ reaction can be
described by adding to this Hamiltonian the electromagnetic
interaction $V_{em}$ introduced in Ref.\ \cite{FdJ_ppg}.
The model Hamiltonian then takes the form
\begin{eqnarray}
H = H_0 + H_{int} + V_{em},
\label{def_hamiltonian}
\end{eqnarray}
where $H_0$ is the sum of relativistic free energy
operators, e.g. $\sqrt{p^2 +m^2_\alpha}$
for the $\alpha = N$, $\Delta$ and $\pi$ degrees of freedom.
We will neglect the nonresonant pionic interactions
($H_2^\prime$ of Eq.(1.2) of Ref.\ \cite{Lee_2}) which are found in
Ref.\ \cite{Lee_2} to be unimportant in describing $NN$ scattering.
The $N\Delta$ direct interaction is also neglected in
Refs.\ \cite{Lee_1,Lee_2} for simplicity. The considered
hadronic interaction is then of the following form
\begin{eqnarray}
H_{int} &= &\sum_{i=1}^{2}[h_{\pi N,\Delta}(i)+h_{\Delta,\pi N}(i)]
\nonumber \\
        &+&\frac{1}{2}\sum_{i,j=1}^{2}[V_{NN,NN}(i,j) + V_{NN,N\Delta}(i,j) +
        V_{N\Delta,NN}(i,j)]
\end{eqnarray}
The derivation of the $NN$ scattering equation from the hadronic Hamiltonian
$H_0 + H_{int}$
can be found in in Ref.\ \cite{Lee_3} and was summarized in
Ref.\ \cite{Lee_2}.
For our present purpose, we will neglect the much weaker effects due to
the $N\Delta$ scattering
(induced by the vertex interaction $h_{\pi N \leftrightarrow \Delta}$) and
the coupling to the $\pi d$ channel. The resulting scattering T-matrix in the
coupled $NN\oplus N\Delta$
space then takes the following form
\begin{eqnarray}
T_{NN,NN}(E) & = & \hat{V}_{NN,NN}(E) +
\hat{V}_{NN,NN}(E)\frac{P_{NN}}{E-H_0 + i\epsilon }T_{NN,NN}(E),
\label{def_Tnn} \\
T_{N\Delta,NN}(E)
   &=& V_{N\Delta,NN}[ 1 + \frac{P_{NN}}{E+H_0 + i \epsilon} T_{NN,NN}(E)], \\
T_{NN,N\Delta}(E) &=& [1 + T_{NN,NN}(E)\frac{P_{NN}}{E-H_0+i\epsilon}]
V_{NN,N\Delta},
\end{eqnarray}
where the effective $NN$ potential is
\begin{eqnarray}
\hat{V}_{NN,NN}(E)= V_{NN,NN} + U^{(1)}_{NN,NN}(E),
\label{def_V_hat}
\end{eqnarray}
with
\begin{eqnarray}
U^{(1)}_{NN,NN}(E) =
V_{NN,N\Delta}\frac{P_{N\Delta}}{E-H_0-\Sigma_{\Delta}(E)}
V_{N\Delta,NN} .
\end{eqnarray}
In the above equations, $P_{NN}$ and $P_{N\Delta}$ are respectively the
projection operators for the $NN$ and $N\Delta$ channels.
The $\Delta$ self-energy $\Sigma_\Delta$
is determined by the $h_{\pi N \leftrightarrow \Delta}$ vertex interaction
in the presence of a spectator nucleon
\begin{eqnarray}
\Sigma_{\Delta}(E) = \sum_{i=1}^{2}h_{\Delta,\pi N}(i) \frac{P_{\pi NN}}
{E - H_0 + i\epsilon}h_{\pi N,\Delta}(i),
\label{def_sig_del}
\end{eqnarray}
where $P_{\pi NN}$ is the projection operator for the $\pi NN$ state.
The vertex interaction $h_{\pi N \leftrightarrow \Delta}$
was determined from fitting the
$\pi N$ $P_{33}$ phase shifts.
The transition potential $V_{NN\leftrightarrow N\Delta}$
was taken from the one-pion-exchange model of Niephaus et al.\ \cite{Niephaus}
with a monopole form factor
$(\Lambda^2-m_{\pi}^2)/(\bar{q}^2+\Lambda^2)$ ($\bar{q}$ denotes the
three-momentum transfer).
The $NN$ interaction is defined by an subtraction of {\em any} $NN$ potential
which fits the $NN$ phase shifts below the pion production threshold.
In this work we consider the Paris potential \cite{Paris} and hence the $NN$
interaction in Eq.\ (\ref{def_V_hat}) is defined by
\begin{eqnarray}
V_{NN,NN} = V_{Paris} - U^{(1)}(E=E_s)
\label{def_Vnn}
\end{eqnarray}
where the substraction energy $E_s$ is a parameter. The above definition of
the $NN$
interaction amounts to removing phenomenologically the two-pion-exchange with
an intermediate
$N\Delta$ state from the Paris potential, in order to avoid the
double counting of the $N\Delta$ effect.
In Refs.\ \cite{Lee_1,Lee_2}, it was found that by choosing
$\Lambda = 650 $ (MeV/c) and $E_s=10$ MeV
(the laboratory energy of the incident nucleon) the solution of
Eq.\ (\ref{def_Tnn}) can best reproduce the phase shifts calculated
from the Paris potential at energies below
about 300 MeV, and also can describe the $NN$ phase shifts reasonably well
up to about 1 GeV.

The numerical method for solving the above coupled-channel equations
Eqs. (\ref{def_Tnn})-(\ref{def_Vnn}) in the momentum-space
representation was well developed in Refs.\ \cite{Lee_1,Lee_2}.
The calculated plane-wave matrix elements
of $T_{NN,NN}(E)$, $T_{NN,N\Delta}(E)$ and $T_{N\Delta,NN}(E)$ are the input
to the study of the $NN\gamma$ reaction.
The formalism presented above yields T-matrices which are a solution
of the non-relativistic Schroedinger equation.
Thus the expression for the $NN$ cross-section calculated with these T-matrices
is
\begin{equation}
\frac{d \sigma}{d \Omega} = \frac{m^2_N}{4 \pi} | T_{NN,NN}(E)|^2,
\label{cross_section}
\end{equation}
where $m_N$ denotes the nucleon mass.

Before we proceed further, it is necessary to define
the electromagnetic interaction $V_{em}$ of Eq.\ (\ref{def_hamiltonian}).
For the $pp \rightarrow pp\gamma$
reaction, it is sufficient to only consider the one-baryon currents since the
leading term of the two-body current is absent in the $pp \gamma$ reaction.
Similar to the approach developed in Ref.\ \cite{Nozawa} to study the
electromagnetic production of pions, the one-baryon current operator
introduced in Ref.\ \cite{FdJ_ppg} is defined by the matrix elements of the
Feynman amplitudes calculated from an effective Lagrangian.
In this way we include the important relativistic spin-correction
into our model.
The matrix-element of the electromagnetic transition potential is
then given by:
\begin{equation}
\langle \bar{p}', \bar{k} \, \lambda | V_{em} | \bar{p} \rangle
=
\sqrt{\frac{m_\alpha}{E^\alpha_{p'}}}
\langle \bar{p}', \bar{k} \, \lambda | \tilde{V}_{em} | \bar{p} \rangle
\sqrt{\frac{m_{\alpha'}}{E^{\alpha'}_p}}.
\end{equation}
Where $\langle \bar{p}' \,\bar{k}, \lambda | \tilde{V}_{em} | \bar{p} \rangle$
denotes the Lorentz-invariant matrix element and
$E^\alpha_p = \sqrt(p^2 + m_\alpha^2)$.
For the $ N \leftrightarrow N \gamma$ vertex this is
($\bar{k} +\bar{p}'= \bar{p}$)
\begin{eqnarray}
\langle \bar{p}^\prime, \bar{k} \, \lambda | \tilde{V}_{em} | \bar{p} \rangle
=
\bar{u}(\bar{p}^\prime)[-i e_i \mbox{$\not \! \epsilon$}
+ \frac{(\mu_i - 1)e} {4m}
(\mbox{$\not \! k$}\mbox{$\not \! \epsilon$} -
\mbox{$\not \! \epsilon$}\mbox{$\not \! k$})]
u(\bar{p})
\label{nuc_vertex}
\end{eqnarray}
where $\lambda$ denotes a photon state with polarization $\epsilon_\mu$.
We follow the conventions of Bjorken and Drell \cite{Bjorken}
The nucleon Dirac spinor is denoted as $u(\bar{p})$, normalized as
$\bar{u}(\bar{p}) u(\bar{p}) = 1$. The charge and anomalous
magnetic moment of the $i$th nucleon are denoted respectively by
$e_i$ and $\mu_i$.
Similarly, the matrix elements of the  $N \rightarrow \gamma \Delta $
and $\Delta \rightarrow \gamma N$ vertex interactions are respectively
\begin{eqnarray}
\langle \bar{p}_{\Delta}, \bar{k} \, \lambda | \tilde{V}_{em} | \bar{p} \rangle
=
\bar{\psi}^{\mu}(\bar{p}_\Delta) \Gamma^{\Delta N\gamma}_{\mu} u(\bar{p})
\end{eqnarray}
and
\begin{eqnarray}
\langle \bar{p}, \bar{k} \, \lambda | \tilde{V}_{em} | \bar{p}_{\Delta} \rangle
=
\bar{u}(\bar{p}) \Gamma^{N \Delta \gamma}_{\mu} \psi^{\mu}(\bar{p}_{\Delta})
\end{eqnarray}
where $\psi^{\mu}(\bar{p}_{\Delta})$ is the Rarita-Schwinger spinor
with normalization $\bar{\psi}^\mu(\bar{p}) \psi_\mu(\bar{p}) = -1$.
We follow Jones and Scadron \cite{Jones} to write the gauge-invariant
vertex functions:
\begin{eqnarray}
\Gamma_\mu^{N \Delta \gamma} &=&
K^1_\mu + K^2_\mu \hspace{5mm}
\label{vertex_1}
\end{eqnarray}
with
\begin{eqnarray}
K^1_\mu &=& ie G_1 (\not \! k \epsilon_\mu - \not \! \epsilon k_\mu)
\gamma^5 T_z
\label{K_1} \\
K^2_\mu &=& ie G_2 (\epsilon_\mu P \cdot k - \epsilon \cdot P k_\mu)
\gamma^5 T_z.
\label{K_2}
\end{eqnarray}
For the decay of a $\Delta$ in a nucleon and photon we have:
\begin{equation}
\Gamma_\mu^{\Delta N \gamma} =
-K^1_\mu + K^2_\mu.
\label{vertex_2}
\end{equation}
In the above expressions $k_\mu = p_\mu^{in} - p_\mu^{out}$
is the photon momentum (defined to be outgoing from the vertex) and
$P = \frac{1}{2}(p_\mu^\Delta + p_\mu^N)$.
$T_z$ is third component of the isospin transition matrix for coupling
an isospin 3/2 to an isospin 1/2 particle.

The coupling constants $G_1$ and $G_2$ in Eqs.\ (\ref{K_1}) and (\ref{K_2})
are conventionally determined by fitting to the $M1^+$ and $E1^+$
multipole data on the photoproduction of pions from
nucleons \cite{Nozawa,Jones,Davidson,Blomqvist,Koch}.
The values obtained depend on the treatment of the non-resonant background
contributions.
Although this leads to some uncertainty in the values, the parameters
found in the literature are not too far apart.
It has also been shown \cite{Blomqvist,Koch} that in order to accurately
reproduce the
$M1^+$  multipole data on the pion photoproduction around the
resonance energy, one needs energy-dependent couplings $G_1$ and $G_2$.
However, given the uncertainty in the coupling constants we ignore
this dependence.
Bearing in mind that the vertex $K_\mu^1$ gives the dominant contribution,
we can classify the various sets of coupling constants by the
magnitude of $G_1$.
The lowest value is found by Nozawa et al.\ \cite{Nozawa}:
$G_1 = 2.024$ (GeV$^{-1}$) and $G_2 = -0.851$ (GeV$^{-2}$).
Highest values are given by Jones and Scadron \cite{Jones}:
$G_1 = 2.68$ (GeV$^{-1}$) and $G_2 = -1.84$ (GeV$^{-2}$) and by
Davidson et al.\ \cite{Davidson}:
$G_1 = 2.556$ (GeV$^{-1}$) and $G_2 = -1.62$ (GeV$^{-2}$).
Note that the authors of Ref. \cite{Davidson} have a slightly different
definition of $K_2$.
An alternative way to extract the $N \Delta \gamma$ coupling parameters is
to assume vector-meson dominance.
On the $N \Delta \gamma$ vertex only the isospin-1 vector meson
contributes and the coupling strengths are determined by the ratio of
$g_{\rho N N}$ and $g_{\rho N \Delta}$. Using the coupling constants of
Ref. \cite{terHaar} this procedure gives
$G_1 = 2.0$ (GeV$^{-1}$) and $G_2 = 0$ (GeV$^{-2}$),
comparable with the values from pion photoproduction.

In the description of $NN$ bremsstrahlung reactions, where the calculations
are based on $NN$-potential models, one makes use of the Lorentz invariant
nature of the various $NN\gamma$ transition amplitudes describing
different bremsstrahlung processes.
It is then important to be able to obtain such invariant amplitudes
from potential models that are based on non-relativistic approaches,
which yield Galilean invariant amplitudes.
This is the case of the present model given by
Eqs.\ (\ref{def_hamiltonian})-(\ref{def_Vnn}).
In Ref.\ \cite{Herrmann1} a procedure is described to construct
Lorentz-invariant amplitudes from non-relativistic amplitudes.
Following this prescription we construct an amplitude that transforms
covariantly by
including the proper $ E/m$ factors and using the proper relativistic
kinematics.
This procedure is similar to the introduction of ``minimal relativity''
in earlier studies of relativistic effects on nuclear bound states
\cite{relativity}.
However, although the procedure provides us with matrix elements
that transform covariantly, our formulation is not fully relativistic.
It suffers from the inconsistency commonly encountered in constructing
realistic meson-exchange $NN$ models.
This deficiency is unavoidable in practice since we do not have a
reliable relativistic theory for describing the short-range interactions
which are clearly beyond the meson-exchange description.
We also point out that our formalism is equivalent to what one
obtains when starting out from a field-theoretic point of view and:
a) approximates all propagators by their positive energy content, and
b) applies a three-dimensional Thompson-like reduction to all integrations
over the four-momentum.

With the ``minimal relativity'' prescription, and in
first order of the electromagnetic coupling, the Lorentz-invariant amplitude
of the $pp \rightarrow pp\gamma$ reaction defined by the model Hamiltonian
Eq.\ (\ref{def_hamiltonian}) is
\begin{eqnarray}
\langle pp\gamma | \tilde{M}(E) | pp \rangle  =
\langle pp\gamma | \tilde{M}^{ext} | pp \rangle
+\langle pp\gamma | \tilde{M}^{resc} | pp \rangle
\label{ppg_total}
\end{eqnarray}
where the external, or single-scattering, term is
\begin{eqnarray}
\langle pp\gamma | \tilde{M}^{ext} | pp \rangle &=&\sum_{\alpha = N,\Delta}
\langle  pp\gamma | \tilde{V}_{em} | \alpha \rangle  G_{\alpha}(E) \langle
\alpha | \tilde{T}_{N\alpha,NN}(E) | pp \rangle
\nonumber \\
&+&\sum_{\alpha = N, \Delta} \langle pp |
\tilde{T}_{NN,N\alpha}(E-E_{\gamma})
| \alpha\rangle
G_{\alpha}(E-E_{\gamma}) \langle  \gamma \alpha | \tilde{V}_{em} | pp \rangle
\label{ppg_single}
\end{eqnarray}
$E_\gamma$ is the photon-energy and we have introduced the $N$ and $\Delta$
propagators ($E^N_p = \sqrt{p^2 + m_N^2}$ etc.)
\begin{eqnarray}
G_{N}(W)&=& \frac{m_N}{E^N_p} \frac{1}{p_0 - E^N_{p} + i\epsilon },
\label{prop_nuc} \\
G_{\Delta}(W) &=& \frac{m_\Delta^{*}}{E^{\Delta*}_{p}}
\frac{1}{p^0 - E^{\Delta}_{p} -
\frac{m_{\Delta}}{E^\Delta_{p}}
\Sigma^{\Delta}(p^0,\bar{p})},
\label{full_prop_del}
\end{eqnarray}
where $p_0 = W - E_{p_1}$, the energy available for propagation in
the presence of the spectator particle, which has momentum $\bar{p}_1$.
$\tilde{T}_{NN,N\alpha}$ denotes the Lorentz invariant scattering amplitude,
constructed from $T_{NN,N\alpha}$ by \cite{Herrmann1}
\begin{equation}
\langle \bar{p}' | \tilde{T}_{NN,N\alpha} | \bar{p} \rangle =
\left( \frac{E^N_p}{m_N} \right)^{1/2}
\left( \frac{E^N_{p'}}{m_N} \right)^{1/4}
\left( \frac{E^{\alpha *}_{p'}}{m^*_\alpha} \right)^{1/4}
\langle \bar{p}' | T_{NN,N\alpha} | \bar{p} \rangle
\end{equation}
The rescattering term is
\begin{eqnarray}
\langle pp \gamma | \tilde{M}^{resc} | pp \rangle &=&
\sum_{\alpha, \beta, \delta = N,\Delta}
\langle pp | \tilde{T}_{NN,\delta \beta}(E - E_\gamma) | \delta \beta \rangle
G_{\delta \beta}(E - E_\gamma)
\nonumber \\
&\times& \langle \gamma \delta | \tilde{V}_{em} | \alpha \rangle
G_{\alpha \beta}(E)
\langle \delta \beta | \tilde{T}_{\alpha \beta, NN}(E) | pp \rangle
\nonumber
\end{eqnarray}
with
\begin{eqnarray}
G_{\delta \beta}(E - E_\gamma)G_{\alpha \beta}(E) &=&
R^\alpha_{p_1} R^\beta_{p_2} R^\delta_{p_3}
\frac{1}{E - E_\gamma - \tilde{E}^\delta_{p_3} - \tilde{E}^\beta_{p_2}}
\frac{1}{E - \tilde{E}^\alpha_{p_1} - \tilde{E}^\beta_{p_2}},
\nonumber \\
R_p^N = \frac{m_N}{E^N_p}, \mbox{\ }
R^\Delta_p &=& \frac{m^*_\Delta}{E^{\Delta *}_p}, \mbox{\ }
\tilde{E}^N_p = E^N_p,  \mbox{\ }
\tilde{E}^\Delta_p =  E^{\Delta}_{p} -
\frac{m_{\Delta}}{E^\Delta_{p}}
\Sigma^{\Delta}(p^0,\bar{p})
\label{ppg_resc}
\end{eqnarray}
In the calculation of Eq.\ \ref{ppg_resc} we neglect the terms were
more than one of the intermediate particles is a $\Delta$.
Thus we do not include diagrams with a two-$\Delta$ propagator
(which is expected to be small due to the mass difference) and the
diagrams with a $\Delta \Delta \gamma$ coupling.

In the above equations $m_\Delta^{*} = 1.23$ GeV is the physical
mass of the $\Delta$, and $m_\Delta$ is the bare mass of the $\Delta$.
Also $E^{\Delta *}_{p} = \sqrt{p^2 + {m_\Delta^*}^2}$
and $E^\Delta_{p} = \sqrt{p^2 + m_\Delta^2}$.
The $\Delta$ propagator we use is consistent with the relativistic extension
of the $\Delta$-propagator used in Ref.\ \cite{terHaar}.
The specific form of $\Sigma^{\Delta}(p_{\Delta}^0,\bar{p}_{\Delta})$
determines the value of
the bare $\Delta$ mass due to the requirement that the $\Delta$ propagator
is resonant at the physical resonance position,
$p_\Delta^0 = m^*_\Delta$.
The form of $\Sigma_\Delta$ is restricted by the $P_{33}$ phase-shift.
However, one finds satisfactory reproductions of the $P_{33}$ phase-shift
with rather different forms of $\Sigma^\Delta$.
For example, the Bransden-Moorhouse parametrization only has an
imaginary part of the
self-energy, implying $m^0_\Delta = m^*_\Delta$.
In the approach of ter Haar and Malfliet \cite{terHaar} one gets
$m^0_\Delta = 1.46$ GeV.
Lee \cite{Lee_1} reports an accurate reproduction of the $P_{33}$
phase-shift over a large energy range for a value $m^0_\Delta = 1.28$ GeV.
As we will show later, the form of $\Sigma^\Delta$ has a definite
influence on the magnitude of the $\Delta$ decay diagrams.

The various processes contributing to the $NN\gamma$ reaction defined by
Eqs.\ (\ref{ppg_single})-(\ref{ppg_resc}) are illustrated in
Fig.\ \ref{diagrams}.
They can be classified as: (1) the contributions from the nucleon current
(Figs.\ \ref{diagrams}a and \ref{diagrams}b), (2) the contributions from
the direct $\Delta$-decay (Fig.\ \ref{diagrams}c), and (3) the contributions
from the $N\Delta$ rescattering (Fig.\ \ref{diagrams}d).
Note that the $NN$ and $N\Delta$ T-matrix in
each diagram is determined by different collision energies since
the outgoing photon shares the total energy available to the system.
This complicates the calculation if we follow the
conventional approach based on the partial-wave decomposition.
Instead, we directly
carry out the calculation of the diagrams with a $N \Delta \gamma$ vertex
in a plane-wave basis.
The plane-wave T-matrices are constructed from the partial-wave
solutions of Eqs.\ (\ref{def_Tnn})-(\ref{def_Vnn}) by including partial
waves up to $J = 9$.
Calculating in the plane-wave basis allows an exact treatment of
the relativistic features such as that of the vertex interactions
defined by Eqs.\ (\ref{nuc_vertex})-(\ref{vertex_2}).

With the ``minimal relativity" prescription described previously, all
matrix-elements of the $pp\gamma$ process displayed in Fig.\ \ref{diagrams}
are Lorentz invariant. We therefore are allowed to calculate each of them
in any convenient frame.
The calculation of the contributions from the nucleon current
(Figs.\ \ref{diagrams}a,b) is identical to that of
Ref.\ \cite{Herrmann1}, except that the $NN$ T-matrix is now
generated from the coupled-channel equations (\ref{def_Tnn})-(\ref{def_Vnn}).
By setting the coupling term $U^{(1)}$ in Eqs.\ (\ref{def_Tnn}) and
(\ref{def_Vnn}) to zero, we reproduce the results of Ref.\ \cite{Herrmann1}.
For the single-scattering $\Delta$-decay diagrams (Fig.\ \ref{diagrams}c),
it is most convenient to do the calculation in
the c.m.-frame of the initial $NN$ state. Explicitly, from
Eq.\ \ref{ppg_single} we have for this diagram
\begin{eqnarray}
\tilde{M}^{ext}(r_1, r_2, r'_1, r'_2) &=&
\sum_{r''}
\langle \bar{p}'', r'_1 | \tilde{\Gamma}^{\Delta N \gamma} |
\bar{p}', r_1'' \rangle
\frac{m^*_\Delta}{E^{\Delta *}_{p'}}
\frac{1}{p^\Delta_0 - E^\Delta_{p'} - \frac{m^0_\Delta}{E^\Delta_{p'}}
\Sigma^\Delta(p^\Delta_0, \bar{p}')} \nonumber \\
&\times&
\langle \bar{p}', r_1'', r'_2 | \tilde{T}_{N \Delta, NN}(s = 4 {E^N_p}^2)
| \bar{p}, r_1, r_2 \rangle.
\nonumber
\end{eqnarray}
with
\begin{eqnarray}
\langle \bar{p}'', r'_1 | \tilde{\Gamma}^{\Delta N \gamma}
| \bar{p}', r_1'' \rangle &=&
\bar{u}_{r_{1'}}(\bar{p}'') \Gamma^{\Delta N \gamma}_\mu
\psi^{\mu *}_{r_1''} (\bar{p}), \nonumber \\
\langle \bar{p}', r_1'', r'_2 | \tilde{T}_{NN,N \Delta}(s)
| \bar{p}, r_1, r_2 \rangle &=&
\bar{\psi}^{\mu *}_{r_1''}(\bar{p}') \bar{u}_{r'_2}(-\bar{p}')
\tilde{T}_\mu^{N \Delta,NN}(s) u_{r_1}(\bar{p}) u_{r_2}(-\bar{p})
\nonumber\\
p_0^\Delta &=& 2E^N_p - E^N_{p'}.
\label{single_delta}
\end{eqnarray}
In these equations $r$ denotes the spin index of the respective spinor,
$\Gamma^{\Delta N \gamma}_\mu$ is defined in Eq.\ (\ref{vertex_2}),
the momenta $\bar{p}$,$\bar{p}'$ and $\bar{p}''$ are determined by the
external kinematics.
A similar expression is found for the second diagram of Fig. \ref{diagrams}c
where the photon is emitted before the strong interaction.

The expressions for the rescattering diagrams with $\Delta$'s are evaluated
in the same fashion.
We see from Eq.\ (\ref{ppg_resc}) that
the calculation involves an integration over the intermediate state.
Again we use the Lorentz invariance of the $pp \gamma$ amplitude and
evaluate the diagram in a suitable frame.
We find for the rescattering diagram where a nucleon is excited
into a $\Delta$ and a real photon
\begin{eqnarray}
&\tilde{M}^{resc}_{N \Delta \gamma}& =
\int \frac{d \bar{p}''}{(2 \pi)^3} \sum_{r''_1, r''_2, r'''_3}
\nonumber \\
& &
\langle \bar{p}' - \frac{\bar{k}}{2}, r'_1, -\bar{p}' -
\frac{\bar{k}}{2}, r'_2 |
\tilde{T}_{NN, \Delta N}
(s = (E_{\bar{p}' - \frac{\bar{k}}{2}}^N + E_{-\bar{p}' -
\frac{\bar{k}}{2}}^N)^2 - \bar{k}^2) |
\bar{p}'' - \bar{k}, r'''_1, -\bar{p}'', r''_2 \rangle
\nonumber \\
& &
\left( \frac{m^2_N}{{E^N_{p''}}^2}
\frac{m^*_\Delta}{E^{\Delta *}_{\bar{p}'' - \bar{k}}} \right)
\left( \frac{1}{2 E^N_p - 2 E^N_{p''}} \right)
\left( \frac{1}{2 E^N_p - k_0 - E^N_{p'} - E^{\Delta *}_{\bar{p}'' - \bar{k}}
-  \frac{m^0_\Delta}{E^\Delta_{\bar{p}'' - \bar{k}}}\Sigma^\Delta(p_0^\Delta,
\bar{p}'' - \bar{k}')}
\right)
\nonumber \\
& &\langle \bar{p}'' - \bar{k}, r_1''' | \tilde{\Gamma}^{N \Delta \gamma}
| \bar{p}'', r_1'' \rangle
\langle \bar{p}'', r_1'', r_2'' | \tilde{T}^{NN,NN}
(s = 4 {E_p^N}^2) | \bar{p}, r_1, r_2
\rangle.
\nonumber \\
&p_0^\Delta& = E^N_p - k_0.
\label{resc_delta}
\end{eqnarray}
In this expression $k$ is the photon momentum.
The notation is similar to Eq.\ (\ref{single_delta}), we specified
explicitly  all the momenta of the $\Delta N$ T-matrix which is not in its
c.m.\ frame.
Due to the presence of the (complex) $\Delta$ self-energy the denominator
with $\Sigma^\Delta$ has no zero in the region of integration.
The other denominator does have a zero,
evaluating the diagram in the c.m.\ frame of the $NN$ T-matrix allows
the pole to be treated with a simple subtraction method.
The other rescattering diagrams with a $\Delta$ intermediate state are
calculated in the same manner.

Finally we point out that the rescattering diagram where the $\Delta$ is the
spectator has an intermediate state which separates in a proton-$\Delta^+$
and a neutron-$\Delta^{++}$ state. The photon couples to the
nucleon, hence this diagram has a contribution depending on the magnetic
moment of the neutron.
This contribution is gauge-invariant, however the
contribution of the proton-proton-photon vertex is not.
Also, since the introduction of $\Delta$ intermediate states allows for
the exchange of charged mesons in the proton-proton T-matrix,
meson-exchange-currents (MEC) will be non-zero.
This all shows that the inclusion of $\Delta$ intermediate states
has implications on the gauge-invariance of the model.

\section{Results}

In this section, we will first present results in the low energy region where
the experimental data at $E_{lab}=280$ MeV are available. We then present our
predictions at $E_{lab}=550$ Mev for the forthcoming experimental test at
COSY. We will also consider the kinematic region where the $\Delta$
dynamics can be best studied.

Compared with the previous works in the low energy region
($E_{lab} \leq$ about 300 MeV),
an important feature of our approach is that we calculate the T-matrices
$T_{NN,NN}$ and $T_{NN\leftrightarrow N\Delta}$ from a coupled-channel
model which was obtained by extending the Paris potential to
include the coupling with the $N\Delta$ and $\pi NN$ states.
As shown in Refs.\ \cite{Lee_1,Lee_2} and briefly discussed in section II,
the constructed coupled-channel model is as
good as the Paris potential in describing the $NN$ data in the low energy
region below the pion production threshold.
In order to illustrate the on- and off-shell differences between teh
$NN$ interaction based on the Paris potential and our coupled channel
approach, we decompose the $NN$ T-matrix in its spin-isospin
compenents \cite{Nakayama}
\begin{eqnarray}
\langle \bar{p}' | T_{NN,NN} | \bar{p} \rangle
&=& \big[ \alpha_1 P_{S=0} + \alpha_2 P_{S=1} +
i \alpha_3(\vec{\sigma}_1 + \vec{\sigma}_2) \cdot \hat{n}
\nonumber \\
&+& \alpha_4 S_{12}(\hat{q}) + \alpha_5 S_{12}(\hat{Q}) +
\alpha_6 S_{12}(\hat{q}, \hat{Q}) \big] P_T.
\label{decompose}
\end{eqnarray}
where
\begin{eqnarray}
\hat{q} &=& \frac{\bar{p} - \bar{p'}}{| \bar{p} - \bar{p'} |},
\hat{Q} = \frac{\bar{p} + \bar{p'}}{| \bar{p} + \bar{p'} |},
\hat{n} = \frac{\bar{p} \times \bar{p'}}{| \bar{p} \times \bar{p'} |},
\nonumber\\
P_{S=0} &=& \frac{1}{4}(1 - \vec{\sigma}_1 \cdot \vec{\sigma}_2),
P_{S=1} = \frac{1}{4}(3 + \vec{\sigma}_1 \cdot \vec{\sigma}_2)
\nonumber \\
S_{12}(\hat{p}) &=&
3 \vec{\sigma}_1 \cdot \hat{p}  \vec{\sigma}_2 \cdot \hat{p}
-  \vec{\sigma}_1 \cdot \vec{\sigma}_2,
\nonumber \\
S_{12}(\hat{p}', \hat{p}) &=& \frac{3}{2}
(\vec{\sigma}_1 \cdot \hat{p}' \vec{\sigma}_2 \cdot \hat{p} +
\vec{\sigma}_1 \cdot \hat{p} \vec{\sigma}_2 \cdot \hat{p}')
- \hat{p}' \cdot \hat{p}  \vec{\sigma}_1 \cdot \vec{\sigma}_2.
\end{eqnarray}
The first two terms in Eq.\ \ref{decompose}, proportional to the spin
projection operator $P_{S = 0}$ and $P_{S = 1}$,  are the central
spin-single (S = 0) and spin-triple (S = 1) components, respectively.
The third term is the spin-orbit component, the fourth and fifth
terms are the usual tensor components.
The last term in Eq.\ \ref{decompose} is the off-shell tensor component
which vanishes identically on-shell as a consequence of time-reversal
invariance.
$P_T$ in Eq.\ \ref{decompose} denotes the total isospin projection operator.
For proton-proton scattering, it projects out onto  the total
isospin $T = 1$ subspace.
In Fig.\ \ref{T_on_shell} we compare the $NN$ on-shell interaction.
Here, following Ref.\ \cite{Herrmann2}, we show the angle-averaged
magnitude of each spin-isospin compenent with total isospin $T = 1$.
The dashed curves are the results for the Paris potential while the
solid curves stand for the coupled-channel model results.
The only notable differences are in the
$P_{S=1}$ channel.
The small differences seen in Fig.\ \ref{T_on_shell} are expected since the
subtraction method defined by Eq.\ (\ref{def_Vnn}) yields a correction
$U^{(1)}(E) - U^{(1)}(E_s)$ to the Paris potential,
which is very small at low energies and becomes significant only as the
collision energy approaches the pion production threshold.

The main difference between the Paris potential and the coupled-channel
model is in the off-shell T-matrices which describe the
wavefunctions in the interaction region.
As an example, we compare in Fig.\ \ref{T_off_shell}
the half-off-shell T-matrices calculated from these two models at
$E_{lab}= 280$ MeV as a function of the off-shell momentum
$p' = |\bar{p}'|$. Significant differences can be observed in most of the
channels, especially at higher off-shell momenta.
However, the $pp \gamma$ reaction at this
low energy only probes the low momentum region
($p' \leq 0.36$ GeV/c)
in which the differences between the two half-off-shell
T-matrices are much smaller.
Moreover, the $pp \gamma$ reaction is insensitive to the
central $P_{S=0}$ channel \cite{Herrmann4},
which shows the largest difference in Fig.\ \ref{T_off_shell}.
We compare in Fig.\ \ref{Ay_on_shell}
the $NN\gamma$ analyzing powers predicted by the Paris
potential (dashed curve) and the coupled-channel model (solid curve), for
proton angles of 12.4 and 14.0 degrees at $E_{L} = 280$ MeV.
To emphasize the farthest off-shell region, we did not include the
rescattering mechanisms (Figs.\ \ref{diagrams}b,d) in this calculation.
Obviously, the off-shell differences shown in Fig.\ \ref{T_off_shell}
do not lead to any dramatic effects on the $NN\gamma$ reaction at
low energies.
We mention that the analyzing power is most sensitive to the tensor and
central spin-triple components of the $NN$ interaction \cite{Herrmann2}.

We now turn to analyzing the effect due to the $\Delta$ mechanism which was
found in Ref.\ \cite{FdJ_ppg} to be significant already at $E_{L} = 280$ MeV.
At certain kinematic conditions, the $\Delta$ mechanism can increase the
cross-section up to 30 \%.
But it can decrease, less substantially,
the cross-section at other kinematic conditions.
Clearly there is a very large interference effect due to the presence
of the $\Delta$ excitation.
The $\Delta$-current, which is almost exclusively magnetic, interferes
very effectively with the magnetic part of the nuclear current.
The latter gives the dominant contribution at this energy.
To see how this arises within our model, we present
in Fig.\ \ref{AandB} the results at $E_{lab} = 280$ MeV
calculated from various combinations of single scattering
mechanisms (Figs.\ \ref{diagrams}a,c).
We first observe that at this low energy the contribution
from the $\Delta$ excitation alone (dot-dot-dash curve) is less than 10\% of
the contribution from the nucleon current (dot-dash curve).
The $\Delta$ contribution depends only very weakly on the photon angle,
while the nucleonic contribution shows a large variation.
Second, we note that the $\Delta$ contribution (Fig.\ \ref{diagrams}c)
consists of two different amplitudes:
(1) the pre-emission amplitude due to
the emission of the proton before the strong interaction takes place, (2)
the post-emission amplitude due to the emission of photon after the $NN$
collision.
The difference between these two amplitudes is mainly in the
$\gamma N \Delta$ vertex. The pre-emission amplitude is determined by
the vertex $\Gamma^{\gamma N\Delta}$,  and the post-emission
amplitude by $\Gamma^{\gamma\Delta N}$.
{}From the expressions Eqs.\ (\ref{vertex_1}) and (\ref{vertex_2}),
we see that the dominant $K^1_{\mu}$ term of the pre-emission vertex
$\Gamma^{\gamma \Delta N}$ has
a minus sign relative to that of the post-emission vertex
$\Gamma^{\gamma N \Delta}$.
Consequently, the pre-emission and post-emission amplitudes tend to
have opposite effects in interfering with the nucleonic contribution.
This is also illustrated in Fig.\ \ref{AandB}. It is seen that
when the pre-emission amplitude is added to the nucleonic amplitude,
the interference effect is destructive at all angles and  yields
the dashed curve in Fig.\ \ref{AandB}.
On the other hand, the interference due to the post-emission of $\Delta$
is constructive.
By further adding this amplitude to the calculation,
the result is shifted from the dashed curve to solid curve. This sensitive
interference between the nucleonic and $\Delta$ contribution provides
an opportunity to test our model of the $\Delta$ excitation.

In calculating the $\Delta$ contributions to the $NN \gamma$ reaction it
is crucial to use a $\Delta$ propagator which is consistent with the
T-matrices used. As discussed in section II, there are other forms
of the $\Delta$ self-energy found in the literature.
All are constrained by fitting the $\pi N$ phase-shifts in the
$P_{33}$ channel, but results from different formulations of
$\pi N$ scattering.
To show how the form of the $\Delta$ self-energy (in casu the $\Delta$
propagator) affects the results, we show in Fig.\ \ref{sig_del} a results
from our calculation (dashed curve) using Eq.\ (\ref{def_sig_del}) to
determine the $\Delta$ self-energy and a results (dotted curve)
using the self-energy of ter Haar and Malfliet \cite{terHaar}.
The contributions of the nucleonic current are represented by the
solid curve.
In all calculations the same T-matrices based on the Paris potential
including $\Delta$ intermediate states were used.
Although both propagators fit the $P_{33}$ phase-shift, the rather
large difference between the results show the need to use a consistent
$\Delta$ propagator. Consequently, all of the results presented hereafter
are obtained using the self-energy of Eq.\ (\ref{def_sig_del}) in
calculating the $\Delta$ propagator Eq.\ (\ref{full_prop_del}).

The inclusion of the $\Delta$ rescattering diagrams (Fig.\ \ref{diagrams}d)
is one of the main new features of this work.
It provides an additional contribution due to the
$\Delta$ excitation. In Fig.\ \ref{del_resc}, we illustrate its
effects at various laboratory energies in changing the calculated
differential cross sections and analyzing powers.
The contribution from the nucleonic currents alone is denoted by the
dotted curves.
By adding the $\Delta$ single scattering mechanism (Fig.\ \ref{diagrams}c),
we obtain the dashed curves.
The solid curves are obtained when the $\Delta$ rescattering mechanisms
(Fig.\ \ref{diagrams}d)
are also included.
At lower energies the effect of the $\Delta$-rescattering on the analyzing
power is comparable to that of the $\Delta$ single-scattering mechanism.
The contribution of the $\Delta$-rescattering to the cross-section is about
one half of the single-scattering mechanism. At higher energies the
rescattering diagrams are less significant.
We have found that the largest $\Delta$ effect is
due to the rescattering diagram where the $\Delta$ is a spectator
(the rightmost diagram in Fig.\ \ref{diagrams}d).
The calculation of this mechanism involves an integration over
two half-off-shell
$T_{NN\leftrightarrow N \Delta}$ matrices and hence is most sensitive to
the short-range behaviour of the transition potential
$V_{NN\leftrightarrow N\Delta}$ potential used in solving the coupled
equations Eqs.\ (\ref{def_Tnn})-(\ref{def_Vnn}).
In the construction of Refs.\ \cite{Lee_1,Lee_2}, the model of
$V_{NN\leftrightarrow N\Delta}$ derived by Niephaus
et al.\ \cite{Niephaus} is used.
Because of the use of a different regularization of the short-range part of
the one-pion exchange, this model is rather different from the model
employed in the $NN\gamma$ calculation of Ref.\ \cite{FdJ_ppg}.
Consequently, the resulting effect of $\Delta$ rescattering is rather
different.
The $\Delta$ rescattering effect calculated using the
$T_{NN\leftrightarrow N \Delta}$ matrices employed in Ref.\ \cite{FdJ_ppg} is
much smaller in changing the differential cross sections.
This is also the reason why the structure of the $\Delta$-contributions
to the analyzing power at 280 MeV in the present work differs from the
results found in Ref.\ \cite{FdJ_ppg}.
We hope that the new high-precision measurements scheduled at the
CELSIUS facility in Uppsala (Sweden) and the COSY ring in Juelich (Germany)
will shed more light on this issue.

In Figs.\ \ref{xsc_280} and \ref{ay_280}, we compare our results with
the experimental data at $E_{lab}=$ 280 MeV.
All mechanisms in Fig.\ \ref{diagrams} are included in the calculations.
The scattering T-matrices and the $\Delta$ self energy needed in
evaluating the $\gamma NN$
amplitudes defined by Eqs.\ (\ref{ppg_total})-(\ref{full_prop_del})
are generated from the model of Refs.\ \cite{Lee_1,Lee_2},
as discussed in section II.
Since there are still some uncertainties in determining from the pion
photoproduction reaction the values of $G_1$ and $G_2$ of the
$\gamma N\Delta$ vertex, we performed calculations for two sets of
coupling constants.
The dashed curves are calculated from using
the highest value of $G_1 = 2.68$ (GeV$^{-1}$) and $G_2 = -1.84$ (GeV$^{-2}$)
as determined by Jones and Scadron \cite{Jones}.
As for their application in $pp \gamma$ calculations these
values are very close to the ones found in a recent analysis of
Lee, who finds slightly higher values \cite{Lee_4}:
$G_1 = 2.89$ (GeV$^{-1}$) and $G_2 = -2.18$ (GeV$^{-2}$).
The solid curves are from using the lowest value of $G_1 = 2.0$ (GeV$^{-1}$),
$G_2 = 0.0$ (GeV$^{-2}$)
as predicted by the vector dominance model.
As a reference we also include the results (dotted curves) calculated
from only the nucleonic contributions
(Figs.\ \ref{diagrams}a,b).
As can be seen in Fig.\ \ref{xsc_280},
the $\Delta$ excitation mechanisms increase the cross-section in most of
the range of kinematics considered.
The $\Delta$-decay diagrams (Fig.\ \ref{diagrams}c) are responsible for the
largest part of the increase.
However, the nucleonic contributions are still dominant at this low energy
and the differences due to the use of two different
sets of $\gamma N \Delta$ coupling constants are not particularly large.
Nevertheless, the $\Delta$ effects clearly significantly bring the
theoretical values closer to the data of the differential cross sections.

The $\Delta$ effects also improve the agreements with the data of
the analyzing powers, as seen in Fig.\ \ref{ay_280}.
The agreement with the data is very good in most of the kinematic
regions considered.
Again, the differences between the results using two
different sets of $G_1$ and $G_2$ are rather small.
Clearly, it is necessary to consider the higher energy regions in order to
have a critical test of our model of the $\Delta$ excitation and pin down
the values of $G_1$ and $G_2$ of the $\gamma N\Delta$ vertex.

To illustrate the increasing importance of the $\Delta$ mechanisms as energy
increases, we show in Fig.\ \ref{inclusive} the inclusive photon
production cross-section as a function of photon energy at photon
angle $\theta=90^0$ and $E_{lab}=280, 550$ and 800 MeV.
At $E_{lab} = 280$ MeV we find a small decrease in the cross-section
for photon energies smaller than 80 MeV and a minor increase
at higher photon energies.
Although the cross sections are clearly dominated by
the nucleonic contribution, the $\Delta$ effect is still significant in
influencing the exclusive cross sections as displayed in
Figs. \ref{xsc_280} and \ref{ay_280}.
At higher laboratory energies we clearly see the $k$-dependence of the
$\Delta$-decay contributions, resulting in peaks at the
high end of the kinematically allowed photon energies.
The cross-section near the $\Delta$ peak is doubled by the $\Delta$ mechanisms
at $E_{lab}=550$ MeV, and by a factor of about 5 at $E_{lab}=800$ MeV.
The large enhancement of the $\Delta$ peak is mainly due to the resonant
behavior of the $\Delta$ propagator Eq.\ (\ref{full_prop_del}) and
the dependence on the photon momentum of the vertex
Eqs.\ (\ref{vertex_1},\ref{vertex_2}).
This is in contrast to the results at $E_{lab} = 280$ where the effect is
due to strong interference of a small $\Delta$-contribution with the
much larger nucleonic current contribution.

In the figure for $E_{lab} = 800$ MeV we also plot the cumulative
contributions of the partial waves for $J \leq 1$ to $J \leq 4$.
The largest contributions are from the $J = 3$ partial wave($^3F_3$).
It is interesting to point out that the same large contribution from the
$J=3$ partial wave is also found in the study of
$pp \rightarrow p n \pi^+$ reaction
(Table I and III of Ref.\ \cite{Matsuyama_2}).
To have a unified description of both the pionic and electromagnetic
excitation in pp scattering, it will be interesting to also have
the $NN\gamma$ data at 800 MeV.

For future experimental tests of our model of the $\Delta$ excitation,
we present our predictions of the exclusive cross sections and
analyzing powers at $E_{lab}=$ 550 MeV (Fig.\ \ref{550}) and
800 MeV (Fig.\ \ref{800}).
The calculations are identical to that of Figs.\ \ref{xsc_280} and
\ref{ay_280}.
The dotted curves are from calculations including only the nucleonic
contributions.
The solid and dashed curves are respectively calculated by
using the $\gamma N\Delta$ coupling constants of vector dominance
and Jones and Scadron \cite{Jones}.
As seen in both Figs.\ \ref{550} and \ref{800},
including the $\Delta$ contributions drastically changes the
differential cross sections.
Hopefully, the forthcoming COSY experiment will have enough accuracy to
distinguish the solid and dotted curves.
This will help narrow down the values of $G_1$ and $G_2$ of the
$\gamma N\Delta$ vertex.
Rather surprisingly, the inclusion of the $\Delta$-decay diagrams has much
less effects on the calculated analyzing powers.
This suggests that the $\Delta$-decay diagrams have at these energies and
kinematics a similar spin-structure (to the extent this is measured by the
analyzing power) as the nucleonic contributions.

Finally, in regard to $pp \gamma$ experiments at high energies we
should mention the efforts of Ref.\ \cite{Nefkens}.
These authors performed a $pp \gamma$ experiment at $E_{lab} = 730 MeV$.
However, due to the experimental set-up they probe a kinematical
region which has relatively low photon-energies.
In the coplanar geometries for which we presented our calculations
the photon-energy in the initial $NN$ c.m. frame ranges from
200 - 350 MeV. In the experiment of Ref.\ \cite{Nefkens},
the photon-energy is below 150 MeV.
The data from Ref.\ \cite{Nefkens} show a rise at the high-energy end of the
kinematically reachable photon-energies, possibly indicating an onset of
the $\Delta$ effect.
However, as can be seen from Fig.\ \ref{inclusive}, the $\Delta$ effect
reaches its maximum at much larger photon energies.
Together with the large errors this makes the experiment unsuitable
for discriminating between the various sets of $N \Delta \gamma$ coupling
constants.

\section{Summary and Discussion}

In this work we have developed an approach to investigate the
$pp$ bremsstralung at both low and intermediate energies.
It is based on an extension of the coupled
$NN\oplus N\Delta\oplus \pi NN$ model of Refs.\ \cite{Lee_1,Lee_2}
to include the electromagnetic coupling with the $N$ and $\Delta$ currents.
The hadronic part of the model is consistent with $NN$ scattering up to 1 GeV.
The $\gamma N\Delta$ coupling is determined
in the study of pion photo-production
in the $P_{33}$ channel.
The relativistic features of the one-baryon current
matrix elements are treated exactly by performing the calculation directly
in momentum space.
In addition, the $\Delta$ rescattering contribution (Fig.\ \ref{diagrams}d)
is evaluated first time in this field.

We have shown that the $\Delta$ contribution can interfere strongly with the
nucleonic contribution.
As seen in Figs. \ref{xsc_280} and \ref{ay_280}, the $\Delta$ contribution
can significantly improve the agreement with the data even at low energies
below pion production threshold.
Given the uncertainty in the normalization of the data, it remains to be
seen if the remaining discrepancies with the data are due to
these normalization problems or that
higher order interaction currents such as those due to the
$\rho \rightarrow \pi \, \gamma$ and $\omega \rightarrow \pi \, \gamma$
couplings can account for these discrepancies.
A qualitative study of these effects in conjuction with the
$\Delta$ effect can be found in Ref.\ \cite{Jetter_2}.

At intermediate energies, the $\Delta$ excitation dominates the
$pp\rightarrow pp\gamma$ cross sections, as displayed in
Figs. \ref{inclusive}-\ref{800}.
Our predictions should be reasonable in guiding the experimental efforts.
However, the employed coupled-channel model was constructed by fitting
the $NN$ phase shift data back in 1981 \cite{Arndt_1}.
This has to be improved since
the phase shift data have been changed significantly recently \cite{Arndt_2}.
Another important step is to extend the present approach to study $np$
bremsstralung at intermediate energies. This requires the inclusion
of exchange currents within the coupled-channel model. Our effort in this
direction will be published elsewhere.

\acknowledgments

This work was supported in part by COSY, KFA-Juelich, grant nr. 41256714.
One of the others (F.d.J) would like to thank the physics division
of Argonne National Laboratory for the hospitality extended to him
during two visits when part of this work was completed.

\figure{
\label{diagrams}
Diagrams included in the calculation (a single line denotes a nucleon, a
double line a
$\Delta$ intermediate state): single-scattering diagrams with
$T_{NN-NN}$ (a), rescattering diagrams with $T_{NN-NN}$ (b),
single-scattering diagrams with $T_{NN-N \Delta}$ (c) and
rescattering diagrams with $T_{NN-N \Delta}$ (d).
}

\figure{
\label{T_on_shell}
Magnitude of the spin-isospin components of the
on-shell interaction,
averaged over the scattering angle as a function of the laboratory energy.
The solid line stands for the T-matrix
including $\Delta$-intermediate states, the dashed line represents the Paris
T-matrix.
}

\figure{
\label{T_off_shell}
Magnitude of the spin-isospin components of the
half off-shell interaction,
averaged over the scattering angle as a function of the off-shell
momentum at a beam energy $E_{lab}$ = 280 MeV.
The solid line stands for the T-matrix
including $\Delta$-intermediate states, the dashed line represents the Paris
T-matrix
}

\figure{
\label{Ay_on_shell}
Coplanar analyzing power for proton-scattering angles $\theta_1$ = 12.4 and
$\theta_2$ = 14.0 as a function of the photon angle, $E_{lab}$ = 280 MeV.
The full line is the result (including the nucleon single scattering diagram,
Fig.\ \ref{diagrams}a) calculated with the T-matrix with $\Delta$
intermediate states, the dashed line is obtained using the Paris T-matrix.
}

\figure{
\label{AandB}
The various contributions to the total cross-section, in a coplanar geometry
with proton-emission angles $\theta_1$ = 27.8 and $\theta_2$ = 28.0,
$E_{lab}$ = 280 MeV.
No rescattering contributions were included.
The dash-dotted line is the result with only the nucleon contributions
(Fig.\ \ref{diagrams}a), the full line stands for the full
model (Fig.\ \ref{diagrams}a,c)
taking the high $N\Delta \gamma$ coupling set
($G_1 = 2.68$ (GeV$^{-1}$), $G_2 = -1.84$ (GeV$^{-2}$)).
The dashed line stands for the nucleonic contributions plus the
pre-emission diagrams.
The dot-dot-dash line is the result with only $\Delta$-decay
diagrams (Fig.\ \ref{diagrams}c).
}

\figure{
\label{sig_del}
The effect of the choice of the $\Delta$ propagator, in a coplanar geometry
with proton-emission angles $\theta_1$ = 27.8 and $\theta_2$ = 28.0,
$E_{lab}$ = 280 MeV.
The dotted line is the result with only the nucleon contributions
(Fig.\ \ref{diagrams}a),
the full line is the result, including all diagrams, calculated with the
$\Delta$ self-energy of ter Haar and Malfliet \cite{terHaar},
the dashed line is the same calculation but using the $\Delta$
self-energy of Eq.\ \ref{def_sig_del}.
In these calculations we took $G_1 = 2.0$ (GeV$^{-1}$) and
$G_2 = 0$ (GeV$^{-2}$).
}

\figure{
\label{del_resc}
The effect of the $\Delta$ rescattering diagrams on both the cross-section
and analyzing power in a coplanar geometry,
at various laboratory energies.
The full line is the result including all diagrams, the dashed line
is calculated with the nucleonic diagrams plus the single-scattering
$\Delta$-decay diagrams (Figs.\ \ref{diagrams}a,b,c) and the dotted line
is calculated with only the nucleonic contributions (Fig. \ref{diagrams}a,b).
In these calculations we took $G_1 = 2.0$ (GeV$^{-1}$) and
$G_2 = 0$ (GeV$^{-2}$).
}

\figure{
\label{xsc_280}
Coplanar geometry $pp\gamma$ cross-section in the laboratory frame as
a function of the photon-emission angle at an incident energy of
$E_{lab}$ = 280 MeV for various proton scattering angles $\theta_1$ and
$\theta_2$.
All diagrams were calculated using the NN and $N\Delta$ T-matrices
of Lee \cite{Lee_1}.
The dotted line corresponds to the calculation which only incorporates
the nucleonic diagrams (Figs.\ \ref{diagrams}a,b).
The results including $\Delta$-decay diagrams are
represented by the solid and the dashed lines.
These correspond to the two choices of the $N \Delta \gamma$ coupling
constants: the full line stands for the results calculated with
$G_1 = 2.0$ (GeV$^{-1}$) and $G_2 = 0$ (GeV$^{-2}$) and the dashed one for
$G_1 = 2.68$ (GeV$^{-1}$) and $G_2 = -1.84$ (GeV$^{-2}$).
The data \cite{Michaelian} do {\em not} contain the arbitrary
normalization factor of 2/3.
}

\figure{
\label{ay_280}
Same as Fig.\ \ref{xsc_280} for the analyzing power.
The data are from Ref.\ \cite{Michaelian} and have been multiplied by a factor
of $-1$.
}

\figure{
\label{inclusive}
Inclusive cross-section at photon-angle $90^\circ$ at various laboratory
energies as a function of photon-momentum.
The dashed line is the result with only the nucleonic contributions
(Fig.\ \ref{diagrams}a,b), the solid line is the full result including
all diagrams calculated with $G_1 = 2.0$ (GeV$^{-1}$) and
$G_2 = 0$ (GeV$^{-2}$).
In the plot for $E_{lab} = 800$ we also show the cumulative contributions
of the partial waves: $J \leq 1$ (dotted), $J \leq 2$ (dash-dot),
$J \leq 3$ (dash-dot-dot) and $J \leq 4$ (short dash).
The full result is calculated with $J \leq 9$.
}

\figure{
\label{550}
Same as Figs.\ \ref{xsc_280} and \ref{ay_280} but at an incident energy of
$E_{lab}$ = 550 MeV.
}

\figure{
\label{800}
Same as Figs.\ \ref{xsc_280} and \ref{ay_280} but at an incident energy of
$E_{lab}$ = 800 MeV.
}


\newpage
\begin{references}

\bibitem{Michaelian} K. Michaelian {\it et al.},
Phys. Rev. D {\bf 41}, 2689 (1990).

\bibitem{Przewoski} B. v. Przewoski, H.O. Meyer, H. Nann, P.V. Pancella,
S.F. Pate, R.E. Pollock, T. Rinckel, M.A. Ross and F. Sperisen,
Phys. Rev. C {\bf 45}, 2001 (1991).

\bibitem{Fearing} R.L. Workman and H.W. Fearing,
Phys. Rev. C {\bf 34}, 780 (1986),
H.W. Fearing, Nucl. Phys. {\bf A463}, 95c (1987).

\bibitem{Herrmann3} V. Herrmann and K. Nakayama,
Phys. Rev. C {\bf 45}, 1450 (1992).

\bibitem{Herrmann1} V. Herrmann and K. Nakayama,
Phys. Rev. C {\bf 46}, 2199 (1992).

\bibitem{Brown} V.R. Brown, P.L. Anthony and J. Franklin,
Phys. Rev. C {\bf 44}, 2199 (1992).

\bibitem{Jetter} M. Jetter, H. Freitag and H.V. von Geramb,
Phys. Scr. {\bf 48}, 229 (1993).

\bibitem{Katsogiannis} A. Katsogiannis and K. Amos,
Phys. Rev. C {\bf 47}, 1376 (1993).

\bibitem{FdJ_ppg} F. de Jong, K. Nakayama, V. Herrmann, and O.Scholten,
Phys. Lett. B. {\bf 333}, 1 (1994).

\bibitem{Lee_1} T.-S. H. Lee,
Phys. Rev. Lett. {\bf 50} 1571 (1983),
Phys. Rev. {\bf C29} 195 (1984).

\bibitem{Lee_2} T.-S. H. Lee and A. Matsuyama,
Phys. Rev. {\bf C36} 1459 (1987).

\bibitem{Bohannon} G.E. Bohannon, L. Heller and R.H. Thompson,
Phys. Rev. {\bf C16}, 284 (1977).

\bibitem{Tiator} L. Tiator, H.J. Weber and D. Drechsel,
Nucl. Phys. {\bf A306}, 468 (1978).

\bibitem{Kamal} A. Szyjewicz and A.N. Kamal, in
NN Interactions-1977, AIP Conf. Proc. vol. 41, edited by
F. Measday, H.W. Fearing and A. Strathdee,
American Institute of Physics, New York (1978), p. 502;
in Few Body Systems and Nuclear Forces I, Proceedings of the Graz meeting,
Springer Lecture Notes in Physics vol. 82,
edited by H. Zingl, M. Haftel and H. Zankel, Berlin (1978), p. 88.

\bibitem{Nifen} H. Nifenecker and J.A. Pinston,
Annu. Rev. Nucl. Part. Sci {\bf 40}, 113 (1990).

\bibitem{Metag} V. Metag,
Ann. d. Phys. {\bf 48}, 121 (1991).

\bibitem{Naudet} C. Naudet {\it et al.},
Phys. Rev. Lett. {\bf 62}, 2652 (1989).

\bibitem{Wolf} Gy. Wolf, W. Cassing and U. Mosel,
Nuc. Phys. {\bf A552}, 549 (1993).

\bibitem{Gacilazo} H. Gacilazo and T. Mizutani,
$\pi NN$ System, World Scientific (1990).

\bibitem{Lee_3} T.-S. H. Lee and A. Matsuyama,
Phys. Rev. {\bf C32} 516 (1985).

\bibitem{Niephaus} G.-H. Niephaus, M. Gari, and B. Sommer,
Phys. Rev. {\bf C20}, 1096 (1976).

\bibitem{Paris} M. Lacombe {\it et al.},
Phys. Rev. {\bf C21}, 861 (1980).

\bibitem{Nozawa} S. Nozawa, B. Blankleider and T.-S. H. Lee,
Nucl. Phys. {\bf A513}, 459 (1990).

\bibitem{Bjorken} J.D. Bjorken and S.D. Drell,
Relativistic Quantum Files, McGraw-Hill, New York, 1965.

\bibitem{Jones} H.F. Jones and M.D. Scadron,
Ann. Phys. {\bf 81}, 1 (1973).

\bibitem{Davidson} R. Davidson, N. Mukhopadyay and R. Wittman,
Phys. Rev. Lett. {\bf 56}, 804 (1986).

\bibitem{Blomqvist} I. Blomqvist and J.M. Laget,
Nucl. Phys. {\bf A280}, 405 (1977).

\bibitem{Koch} J.H. Koch, E.J. Moniz, N. Ohtsuka,
Ann. Phys. {\bf 154}, 99 (1984).

\bibitem{terHaar} B. ter Haar and R. Malfliet,
Phys. Rep. {\bf 149}, 207 (1987).

\bibitem{relativity} G.E. Brown and A.D. Jackson,
The Nucleon-Nucleon Interaction, North-Holland, Amsterdam (1976).

\bibitem{Nakayama} K. Nakayama, S. Krewald and J. Speth,
Nucl. Phys. {\bf A451}, 243 (1986).

\bibitem{Herrmann2} V. Herrmann, K. Nakayama, O. Scholten, and H. Arellano,
To appear in Nucl. Phys. A.

\bibitem{Herrmann4} V. Herrmann and K. Nakayama,
Phys. Lett. B {\bf 333}, 251 (1990).

\bibitem{Lee_4} T.-S. H. Lee, in preparation.

\bibitem{Matsuyama_2} A. Matsuyama and T.-S. H. Lee,
Phys. Rev. {\bf C34}, 1900 (1986).

\bibitem{Nefkens} B.M.K. Nefkens, O.R. Sander, and D.I. Sober,
Phys. Rev. Lett. {\bf 38} 876 (1977).

\bibitem{Jetter_2} M. Jetter and H.W. Fearing,
Triumf preprint TRI-PP-94-90.

\bibitem{Arndt_1} R.A. Arndt and L.D. Roper,
Phys. Rev. {\bf D 25}, 2011 (1982).

\bibitem{Arndt_2} SAID interactive dial-in system,
R.A. Arndt, private communication.

\end{references}
\end{document}